\documentclass[
    ,final            
  ]
  {aipproc}

\layoutstyle{6x9}

\graphicspath{{fig/}}

\newcommand{\ie}{{\em i.e.}}

\newcommand{\GeV}{\;\mathrm{GeV}}
\newcommand{\MeV}{\;\mathrm{MeV}}
\newcommand{\nutev}{NuTeV}

\newenvironment{rightcaption}{
\footnotesize
\setlength{\parindent}{0pt}
\addtocounter{figure}{1}
{\bf FIGURE \thefigure.}\hspace*{2pt}
}

\begin{document}


\title{Quark Asymmetries in Nucleons
\footnote{Talk given at DIS 2005, Madison, USA, April 27-May 1, 2005}}


\author{Johan Alwall}{
  address={High Energy Physics, Uppsala University, Box 535, S-75121 Uppsala, Sweden}
}

\begin{abstract}
We have developed a physical model for the non-perturbative $x$-shape
of parton density functions in the proton, based on Gaussian
fluctuations in momenta, and quantum fluctuations of the proton into
meson-baryon pairs. The model describes the proton structure function
and gives a natural explanation of observed quark asymmetries, such as
the difference between the anti-up and anti-down sea quark
distributions and between the up and down valence distributions. We
also find an asymmetry in the momentum distribution of strange and
anti-strange quarks in the nucleon, large enough to reduce the NuTeV
anomaly to a level which does not give a significant indication of
physics beyond the standard model.
\end{abstract}

\maketitle



The low-scale parton density functions give a description of the
hadron at a non-perturbative level. The conventional approach to these
functions is to make parameterizations using some more or less arbitrary
functional forms, based on data from deep inelastic scattering and
hadron collision experiments. Another approach, however, is to start
from some ideas of the behavior of partons in the non-perturbative
hadron, and build a model based on that behavior. The advantage with
this approach is that the successes and failures of such a model
allows us to get insight into the non-perturbative QCD dynamics.  The
model presented here, and described in detail in
\cite{Alwall:2005xd,Alwall:2004rd}, describes the $F_2$ structure
function of the proton, as well as sea quark asymmetries of the
nucleon. Most noteworthy, our model predicts an asymmetry between the
momentum distributions of strange and anti-strange quarks in the
nucleon of the same order as the newly reported results from NuTeV
\cite{Mason}.

\begin{figure}
\includegraphics*[width=30mm]{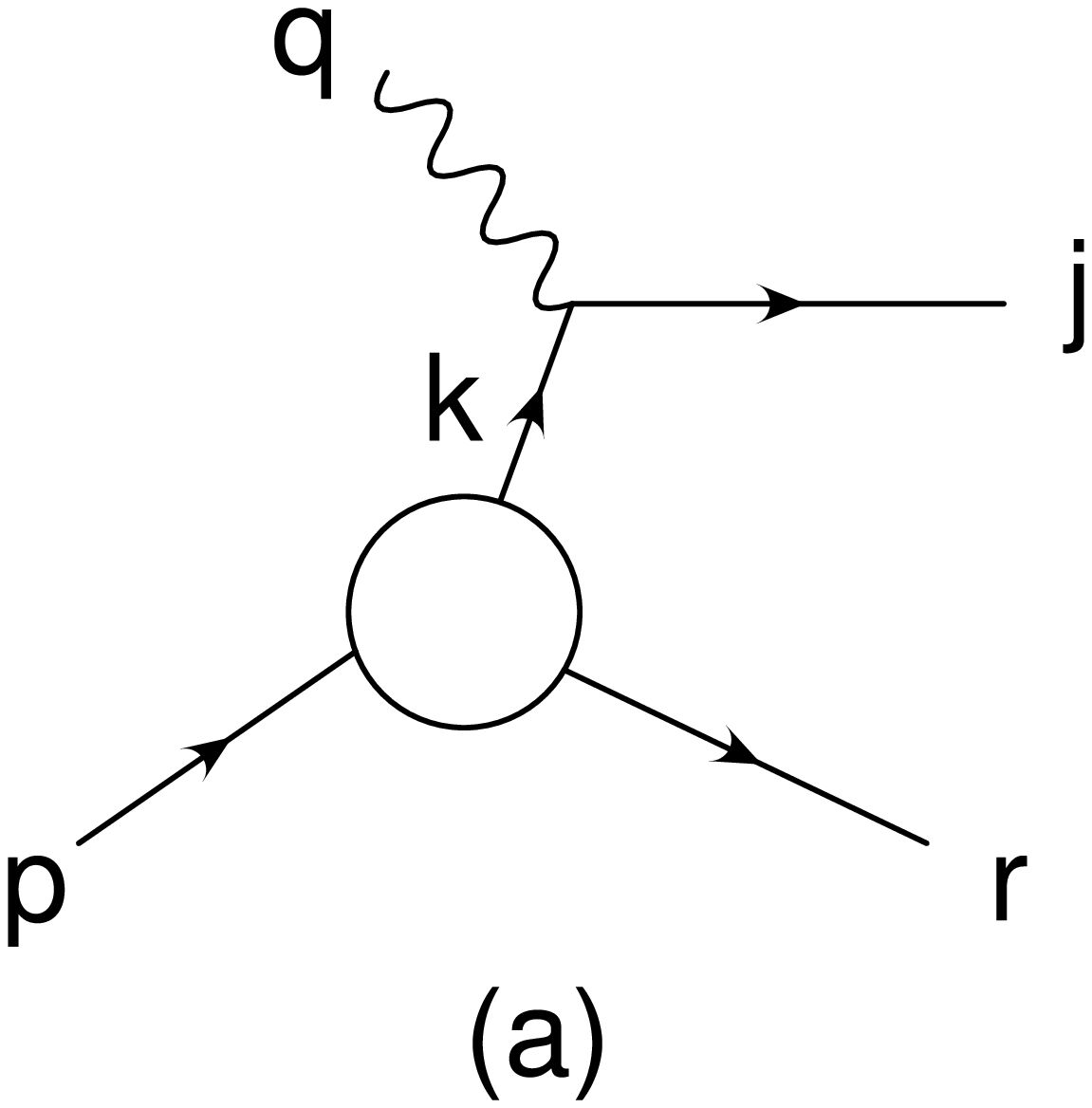}
\includegraphics*[width=30mm]{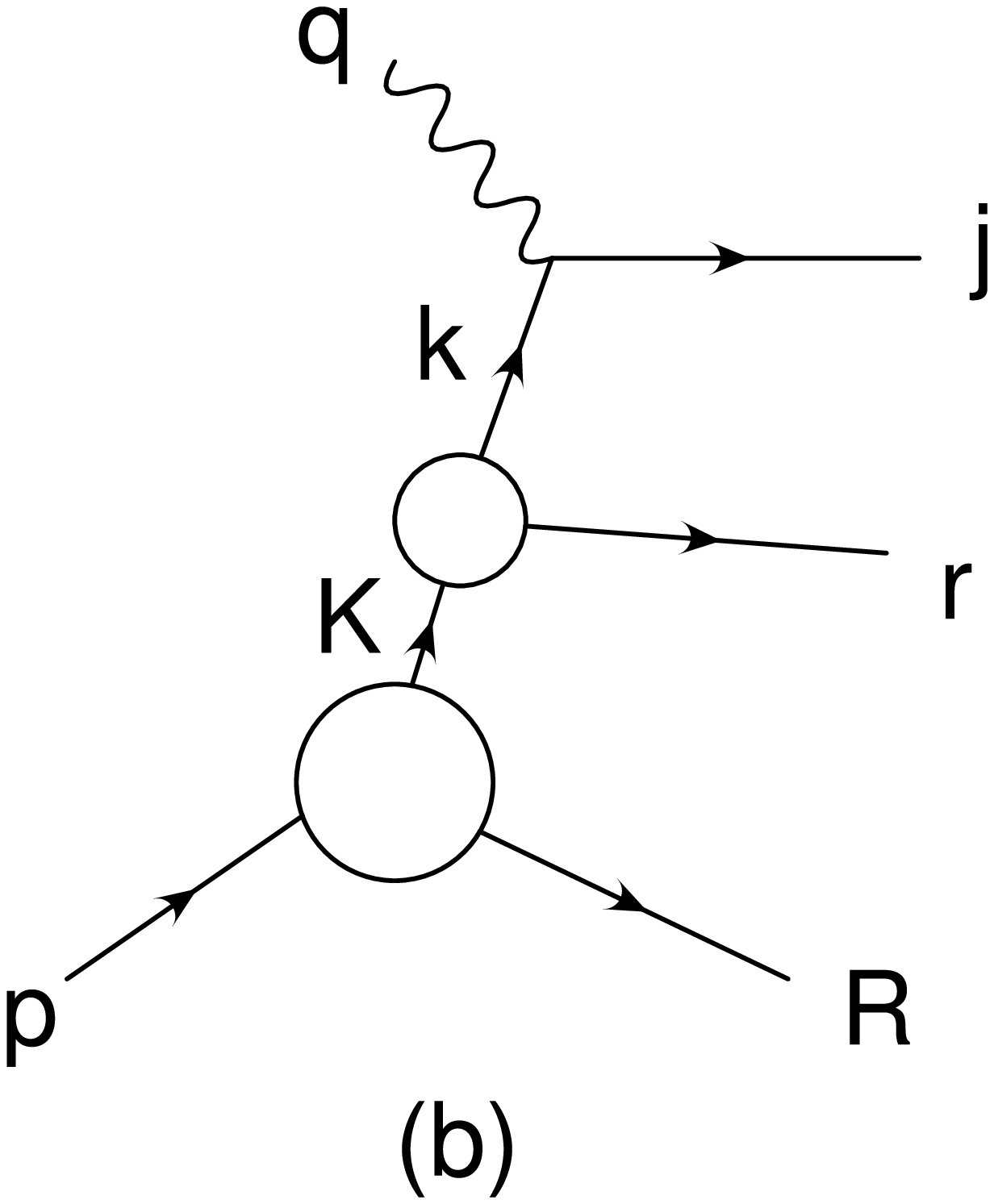}
\includegraphics*[width=50mm,trim=0 15 0 0]{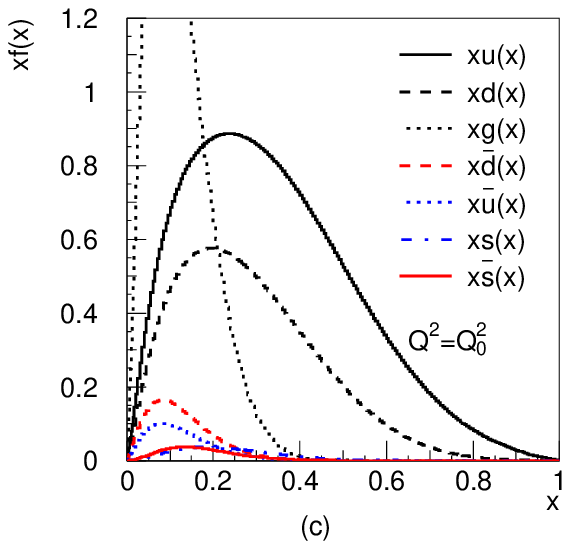}
\caption{\label{fig:fluct} Illustration of the processes (a) probing a
valence parton in the proton and (b) a sea parton in a hadronic
fluctuation (letters are four-momenta). (c) shows the resulting parton
distributions at the starting scale $Q_0^2$.}
\end{figure}

This work extends the model previously presented in
\cite{Edin:1998dz}. The model gives the four-momentum $k$ of a single
probed valence parton (see Fig.~\ref{fig:fluct}a for definitions of
momenta) by assuming that, in the nucleon rest frame, the shape of the
momentum distribution for a parton of type $i$ and mass $m_i$ can be
taken as a Gaussian
$f_i(k) = N(\sigma_i,m_i) \exp\left\{-\left[(k_0-m_i)^2+
k_x^2+k_y^2+k_z^2\right]/2\sigma_i^2\right\}
$
which may be motivated as a result of the many interactions binding
the parton in the nucleon. The width of the distribution should be of
order hundred MeV from the Heisenberg uncertainty relation applied to
the nucleon size, \ie\ $\sigma_i=1/d_N$. The momentum fraction $x$ of
the parton is then defined as the light-cone fraction $x=k_+/p_+$. We
impose constraints on the final-state momenta in order to obtain a
kinematically allowed final state, which also ensures that
$0<x<1$. Using a Monte Carlo method these parton distributions are
integrated numerically without approximations.

To describe the dynamics of the sea partons, we note that the
appropriate basis for the non-perturbative dynamics of the bound state
nucleon should be hadronic. Therefore we consider hadronic
fluctuations, for the proton
\begin{equation} \label{eq:hadronfluctuation}
|p\rangle = \alpha_0|p_0\rangle + \alpha_{p\pi^0}|p\pi^0\rangle +
\alpha_{n\pi^+}|n\pi^+\rangle + \ldots + \alpha_{\Lambda K}|\Lambda
K^+\rangle + \ldots
\end{equation}
Probing a parton $i$ in a hadron $H$ of a baryon-meson fluctuation
$|BM\rangle$ (see Fig.~\ref{fig:fluct}b) gives a sea parton with
light-cone fraction $x=x_H\, x_i$ of the target proton. The momentum
of the probed hadron is given by a similar Gaussian, but with a
separate width parameter $\sigma_H$. Also here, kinematic constraints
ensure that we get a physically allowed final state. The procedure
gives $x_H\sim M_H/(M_B+M_M)$, \ie\ the heavier baryon gets a harder
spectrum than the lighter meson. The normalization of the sea
distributions is given by the amplitude coefficients $\alpha_{BM}^2$ of
Eq.~\eqref{eq:hadronfluctuation}. These cannot be calculated from
first principles in QCD and are therefore taken as free parameters to
be fitted using experimental data.

The resulting valence and sea parton $x$-distributions apply at a low
scale $Q_0^2$, and the distributions at higher $Q^2$ are obtained
using perturbative QCD evolution at next-to-leading order.

The model has in total four shape parameters and three normalization
parameters, plus the starting scale, to determine the parton densities
$u$, $d$, $g$, $\bar{u}$, $\bar{d}$, $s$, $\bar{s}$. These are (with values
resulting from fits to experimental data as described below):
\begin{equation}
\label{eq:params}
\begin{array}{c}
\sigma_u=230\MeV \quad \sigma_d=170\MeV \quad 
\sigma_g=77\MeV \quad \sigma_H=100\MeV\\
\alpha_\mathrm{p\pi^0}^2=0.45 \quad \alpha_{n\pi^+}^2=0.14 \quad 
\alpha_{\Lambda K}^2=0.05 \quad Q_0=0.75\GeV
\end{array}
\end{equation}
The resulting parton densities are shown in Fig.~\ref{fig:fluct}(c).

\begin{figure}
\includegraphics*[width=10cm]{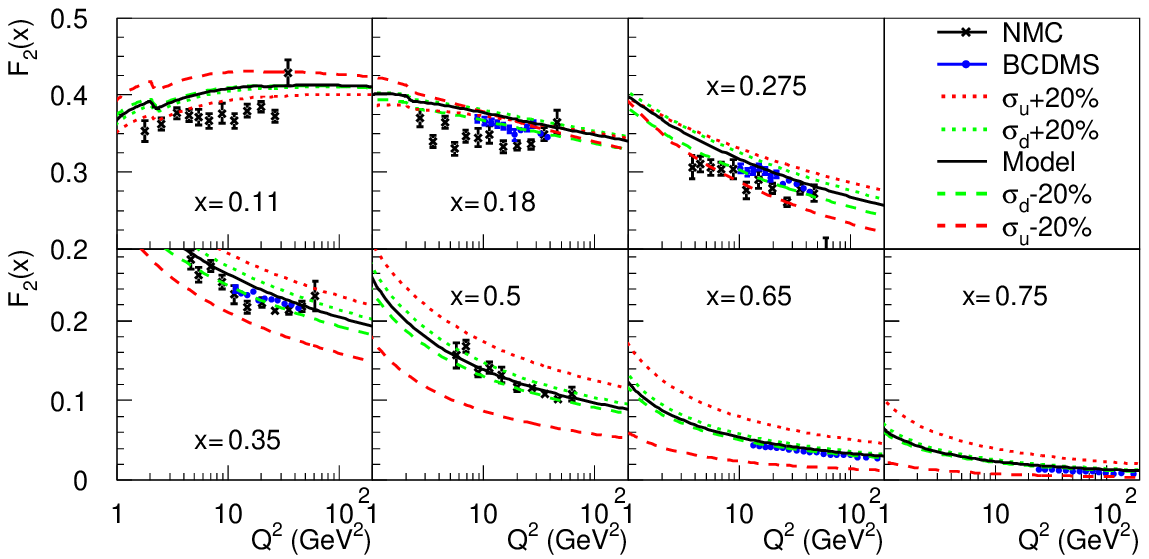}
\includegraphics*[width=5cm]{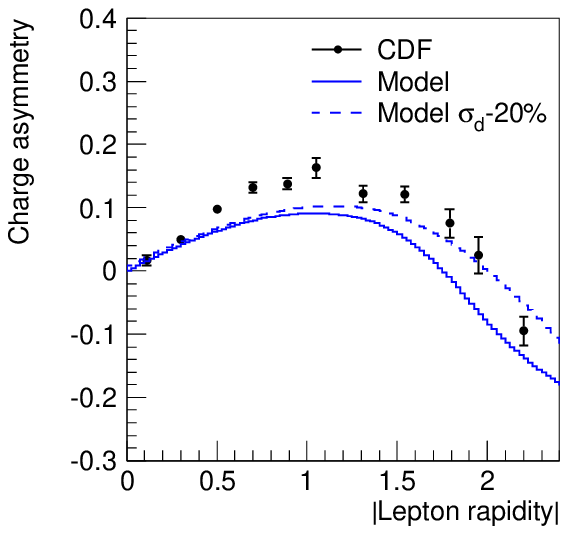}
\caption{\label{fig:xbinned} Left: The proton structure function
$F_2(x,Q^2)$ for large $x$ values; NMC and BCDMS data \cite{NMC,BCDMS}
compared to our model, also showing the results of $\pm 20\%$
variations of the width parameters $\sigma_u$ and $\sigma_d$ for the
$u$ and $d$ valence distributions. Right: The charge asymmetry for
leptons from $W^\pm$-decays in $p\bar p$ collisions at the Tevatron
\cite{w-asym} compared to our model, with best-fit parameters and a
$20\%$ reduced width of the valence $d$ quark distribution.}
\end{figure}

In order to fix the values of the model parameters, we make a global
fit using several experimental data sets: Fixed-target $F_2$ data to
fix large-$x$ (valence) distributions (Fig.~\ref{fig:xbinned}a); HERA
$F_2$ data for the gluon distribution width and the starting scale
$Q_0^2$; $\bar d / \bar u$-asymmetry data for the normalizations of
the $|p\pi^0\rangle$ and $|n\pi^+\rangle$ fluctuations (Fig.~3); and
strange sea data to fix the normalization of fluctuations including
strange quarks (Fig.~4a). We have also compared with $W^\pm$ charge
asymmetry data as a cross-check on the ratio of Gaussian widths for
the $u$ and $d$ valence quark distributions
(Fig.~\ref{fig:xbinned}b). It is interesting to note that this simple
model can describe such a wealth of different data with just one or
two parameters per data set.

In our model, the shape difference between the valence $u$ and $d$
distributions in the proton, apparent from the $W^\pm$ charge
asymmetry data, is described as different Gaussian widths. This would
correspond to a larger effective volume in the proton for $d$ quarks
than for $u$ quarks, an effect which could conceivably be explained by
Pauli blocking of the $u$ quarks.

Since the proton can fluctuate to $\pi^0$ and $\pi^+$ by
$|p\pi^0\rangle$ and $|n\pi^+\rangle$, but to $\pi^-$ only by the
heavier $|\Delta^{++}\pi^-\rangle$, we get an excess of $\bar d$ over
$\bar u$ in the proton sea. Interestingly, the fit to data improves
when we use a larger effective pion mass of 400 MeV (see Fig.~3). This
might indicate that we have a surprisingly large coupling to heavier
$\rho$ mesons, or that one should use a more generic meson mass rather
than the very light pion.

\begin{figure}[b]
\hspace*{-7mm}\begin{minipage}{0.6\columnwidth}
\includegraphics*[width=1\columnwidth,trim=0 20 0 0]{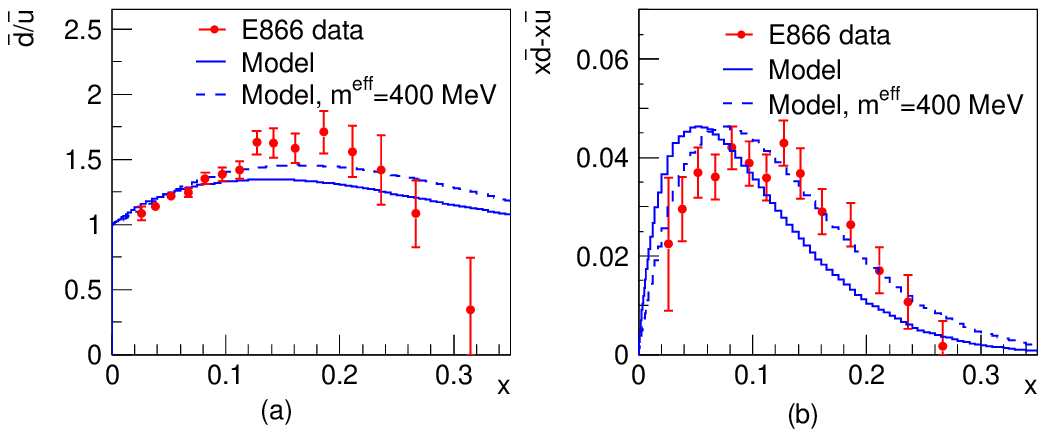}
\end{minipage}
\hspace*{5mm}\begin{minipage}{0.4\columnwidth}
\begin{rightcaption}
Comparison between our model and data from the E866/NuSea
collaboration \cite{dubar}: (a) $\bar u(x)/\bar d(x)$ (b)
$xd(x)-xu(x)$. The full line uses the physical pion mass, while the
dashed line uses an effective pions mass $m^\mathrm{eff} = 400\MeV$ as
discussed in the text.
\end{rightcaption}
\end{minipage}
\end{figure}

The lightest strange fluctuation is $|\Lambda K^+\rangle$. If we let
this implicitly include also heavier strange meson-baryon
fluctuations, we can fit the normalization $\alpha_{\Lambda K}^2$ to
strange sea data (see Fig.~4a). The result corresponds to
$\int_0^1(xs+x\bar s)dx/\int_0^1(x\bar u+x\bar d)dx \approx 0.5$, in
agreement with standard parton density parameterizations. We note that
this indicates a normalization $\propto 1/\Delta
M_{BM}=1/(M_B+M_M-M_p)$ rather than the expected $\propto 1/\Delta
M_{BM}^2$.

Since the $s$ quark is in the heavier baryon $\Lambda$ and the $\bar
s$ quark is in the lighter meson $K^+$, we get a non-zero asymmetry
$S^- = \int_0^1 dx [xs(x)-x\bar s(x)]$ in the momentum distribution of
the strange sea, as seen in Fig.~4b and 5. Depending on details of the
model, we get the range $ 0.0010\leq S^- \leq 0.0023$ for this
asymmetry. 

This is especially interesting in connection to the \nutev\
anomaly \cite{nutev}. \nutev\ found, based on the observable 
$R^- = \frac{\sigma(\nu_{\mu}N\to \nu_{\mu}X)-
            \sigma(\bar{\nu}_{\mu}N\to \bar{\nu}_{\mu}X)}
	   {\sigma(\nu_{\mu}N\to \mu^- X)-
            \sigma(\bar{\nu}_{\mu}N\to \mu^+X)}
   = g_L^2 - g_R^2 = \frac{1}{2} - \sin^2\theta_W$
a $3\sigma$ deviation of $\sin^2\theta_W$ compared to the Standard
Model fit: $\sin^2\theta_W^\mathrm{NuTeV}=0.2277\pm 0.0016$ compared
to $\sin^2\theta_W^\mathrm{SM} = 0.2227\pm 0.0004$.  However, an
asymmetric strange sea would change their result, since $\nu$ only
have charged current interactions with $s$ and $\bar \nu$ with $\bar
s$. Using the folding function provided by \nutev\ to account for their
analysis, the $s$-$\bar s$ asymmetry from our model gives a shift
$-0.0024\leq\Delta\sin^2\theta_W=\int_0^1 dx\, s^-(x)
F(x)\leq-0.00097$, \ie\ the discrepancy with the Standard Model result
is reduced to between $1.6\sigma$ and $2.4\sigma$, leaving no strong
hint of physics beyond the Standard Model.

\begin{figure}
\includegraphics*[width=9cm,trim=0 10 0 0]{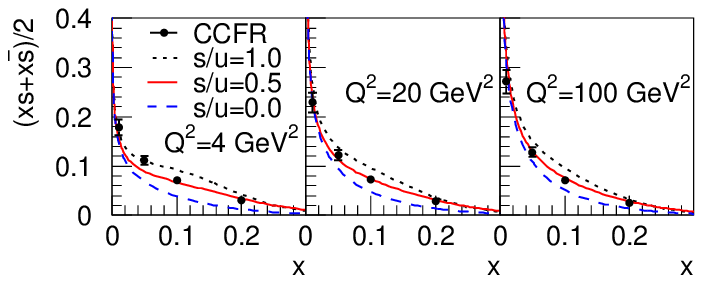}
\includegraphics*[width=5.8cm,trim=0 0 0 10]{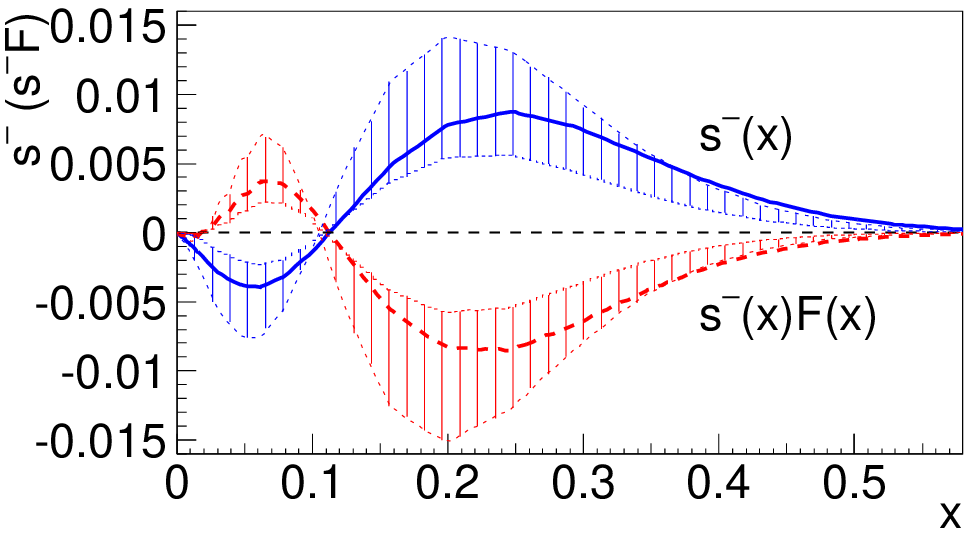}
\caption{\label{fig:sdata}Left: CCFR data \cite{ssbar} on the strange
sea distribution $(xs(x)+x\bar{s}(x))/2$ compared to our model based
on $|\Lambda K\rangle$ fluctuations with different normalizations.
Right: The strange sea asymmetry $s^-(x) = xs(x)-x\bar{s}(x)$ (at
$Q^2=20\GeV^2$) from the model and combined with the function $F(x)$
accounting for NuTeV's analysis giving $\Delta \sin^2\theta_W =
\int_0^1 dx\, s^-(x) F(x) = -0.0017$. The uncertainty bands correspond
to the uncertainties for $S^-$ and $\Delta \sin^2\theta_W$ quoted in the
text.\vspace{-2mm}}
\end{figure}

We have also considered charmed fluctuations. The lightest charmed
baryon-meson fluctuation $|\Lambda_C \overline D\rangle$ gives $c$ and
$\bar c$ distributions as in Fig.~5, where the normalization
$\alpha_{\Lambda_C D}^2$ is taken to be $\propto 1/\Delta M_{BM}$, as
suggested by the strange sea normalization. However, in order to
conform to the EMC $F_2^c$ data at large $x$, a normalization close to
$1/\Delta M_{\Lambda_C \overline D}^2$ seems to be enough
\cite{Alwall}.

\begin{figure}
\hspace*{-7mm}\begin{minipage}{0.45\columnwidth}
\includegraphics[width=1\columnwidth, bb=0 15 156 65]{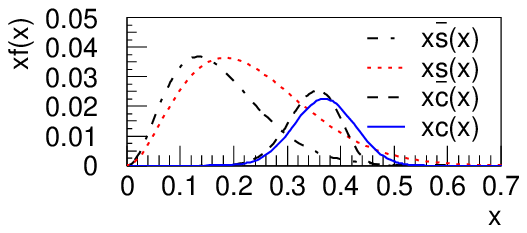}
\end{minipage}
\hspace*{5mm}\begin{minipage}{0.55\columnwidth}
\begin{rightcaption}
Comparison between the strange and charm sea obtained
from our model with the inclusion of the $\Lambda_C\overline D$
fluctuation. The normalization is here taken to be $\propto
1/(M_B+M_M-M_p)$ as suggested by strange sea data.
\end{rightcaption}
\end{minipage}
\end{figure}

{\bf Acknowledgments: } We would like to thank the organizers for
the opportunity to talk at DIS'05, and Stan Brodsky for very interesting
discussions.\vspace{-4mm}







\end{document}